\documentclass[prl,twocolumn,letterpaper,superscriptaddress]{revtex4}
\usepackage{graphicx,amsmath,amssymb,amsbsy,amsfonts,latexsym,color,dcolumn,bm,epsfig,subfigure}
\renewcommand{\imath}[0]{\mathrm{i}}
\newcommand{\mathbfh}[1]{\hat{\mathbf{#1}}}
\newcommand{\abs}[1]{\left\vert#1\right\vert}

\begin{document}

\title{Quantum Rolling Friction}

\author{F. Intravaia}
\affiliation{Humboldt-Universit\"at zu Berlin, Institut f\"ur Physik, AG Theoretische Optik \& Photonik, 12489 Berlin, Germany}
\author{M. Oelschl\"{a}ger}
\affiliation{Max-Born-Institut, 12489 Berlin, Germany}
\author{D. Reiche}
\affiliation{Max-Born-Institut, 12489 Berlin, Germany}
\author{D. A. R. Dalvit}
\affiliation{Theoretical Division, MS B213, Los Alamos National Laboratory, Los Alamos, New Mexico 
87545, USA}
\author{K. Busch}
\affiliation{Humboldt-Universit\"at zu Berlin, Institut f\"ur Physik, AG Theoretische Optik \& Photonik, 12489 Berlin, Germany}
\affiliation{Max-Born-Institut, 12489 Berlin, Germany}

\begin{abstract} 
An atom moving in a vacuum at constant velocity and parallel to a surface experiences 
a frictional force induced by the dissipative interaction with the quantum fluctuations of 
the electromagnetic field. We show that the combination of nonequilibrium dynamics, 
anomalous Doppler effect and spin-momentum locking of light mediates an intriguing 
interplay between the atom's translational and rotational motion. In turn, this  
deeply affects the drag force in a way that is reminiscent of classical rolling friction. 
Our fully non-Markovian and nonequilibrium description reveals counterintuitive features 
characterizing the atom's velocity-dependent rotational dynamics. These results prompt 
interesting directions for tuning the interaction and for investigating nonequilibrium 
dynamics as well as the properties of confined light. 
\end{abstract}

\maketitle

Quantum light-matter interactions continue to fascinate with intriguing and 
non-intuitive phenomena. During the last years, many interesting results 
involving nonequilibrium physics and light confinement in photonic and plasmonic 
systems have been reported. 
Although systems out of equilibrium are very common in nature, only recently 
have intense investigations started to unravel their relevance for both fundamental 
and applied research 
\cite{Esposito09,Polkovnikov11}. 
On the other hand, light confinement is already known for inducing several 
important behaviors. Nonetheless, it continues to surprise and is currently 
attracting attention, for instance, for conveying spin-orbit interactions of 
light (a.k.a. spin-momentum locking)
\cite{Lodahl17,Aiello15,Bliokh15}.
Here, we combine these fields of research within a larger framework: We show that, 
when an atom is forced to move parallel to a surface, \emph{quantum rolling 
frictional dynamics} results from the nonequilibrium interplay of the atomic 
translational and rotational motion. Despite the apparent resemblance to the 
behavior of a classical body rolling on a surface, the underlying physics of 
this phenomenon features many interesting counterintuitive aspects.

Because of quantum fluctuations, light-matter interactions lead to the occurrence 
of non-conservative (frictional) forces on electrically neutral and non-magnetic 
objects 
\cite{Dedkov02a,Volokitin02}. 
These forces are quantum in nature and the physics behind quantum friction is 
related to the quantum Cherenkov effect through the anomalous Doppler effect 
\cite{Frolov86,Nezlin76,Ginzburg96,Maghrebi13a}. 
In this process, real photons are extracted from the vacuum at the cost of 
the object's kinetic energy; they are absorbed and re-emitted thus producing 
a fluctuating momentum recoil 
\cite{Intravaia15}.
When only the atomic translational motion is considered, spin-zero photons are 
absorbed and re-emitted, and a net quantum frictional force that opposes the 
translational motion appears. This anisotropic process was investigated in many 
scenarios during the last decade 
\cite{Dedkov02a,Barton10b,Silveirinha12,Pieplow13,Jentschura15,Klatt16,Volokitin17,Lombardo17,Rodriguez-Lopez18} 
and its connection to nonequilibrium physics was recently highlighted 
\cite{Intravaia16}. 
In this Letter we show that, when the rotational degrees of freedom are 
involved in the dynamics, the atom can also exchange angular momentum, 
absorbing and emitting photons with nonzero spin. However, due to nonequilibrium 
physics, the anomalous Doppler effect and the spin-momentum locking of light 
\cite{Lodahl17,Aiello15,Bliokh15}, 
this stochastic process is peculiarly unbalanced: 
A force \emph{in the direction of the motion} appears and partially compensates 
the translational friction. As a result, a net atomic rotation emerges with 
a sense \emph{opposite} to that of classical rolling.

We consider a system at zero temperature consisting of an atom propelled at 
constant height $z_{a}>0$ parallel to a flat surface at $z=0$.
We focus on the \emph{nonequilibrium steady state} (NESS) characterized by a 
constant velocity $\mathbf{v}$ reached by the system when friction balances 
the external drive. 
In the NESS the frictional force can be written as 
$\mathbf{F}=\mathbf{F}^{\rm t}+\mathbf{F}^{\rm r}$ \cite{Intravaia16a}, 
where
\begin{subequations}
\label{friction2total}
\begin{gather}
\mathbf{F}^{\rm t}=-
	2\int_{0}^{\infty}\hspace{-.3cm}d\omega\int\frac{d^{2}\mathbf{k}}{(2\pi)^{2}} \, \mathbf{k}
		\mathrm{Tr}\left[\underline{S}^{\sf T}_{R}(-\omega^{-}_{\mathbf{k}},\mathbf{v})
								\cdot \underline{G}^{\rm s}_{I}(\mathbf{k},z_{a},\omega)\right],
\label{friction2t}
\\
\mathbf{F}^{\rm r}=
	-2\int_{0}^{\infty}\hspace{-.3cm}d\omega\int\frac{d^{2}\mathbf{k}}{(2\pi)^{2}} \, \mathbf{k}
		\mathrm{Tr}\left[\underline{S}_{I}^{\sf T}(-\omega^{-}_{\mathbf{k}},\mathbf{v})
								\cdot \underline{G}^{\rm as}_{R}(\mathbf{k},z_{a}, \omega)\right].
\label{friction2r}
\end{gather} 
\end{subequations}
$\omega^{\pm}_{\mathbf{k}}=\omega\pm\mathbf{k}\cdot \mathbf{v}$ is the Doppler-shifted 
frequency of the vacuum field in the atom's comoving frame, $\mathbf{k}$ is the 
component of the wave vector parallel to the surface and $\underline{G}(\mathbf{k},z_a,\omega)$ 
is the Fourier transform of the electromagnetic Green tensor 
\cite{Wylie85,Intravaia16a}. 
$\underline{S}(\omega,\mathbf{v})$
is the velocity-dependent atomic power spectrum, i.e. the Fourier transform of the 
stationary two-time correlation tensor
$\underline{C}_{ij}(\tau, \mathbf{v}) =\langle\hat{d}_i(\tau) \hat{d}_j(0) \rangle$ ($i,j=x,y,z$). 
Here, $\mathbfh{d}(t)$ describes the full nonequilibrium quantum dynamics of the 
particle's electric dipole vector operator. The symbol ``$\mathrm{Tr}$'' traces 
over the cartesian indices, while ``${\sf T}$'' stands for the transpose. 
The superscripts ``${\rm s}$'' and ``${\rm as}$'', and the subscripts ``$R$'' and ``$I$''  
indicate the symmetric and the antisymmetric part of tensors and the real and the 
imaginary part of the corresponding quantity, respectively.  

\begin{figure}
\includegraphics[width=0.42\textwidth]{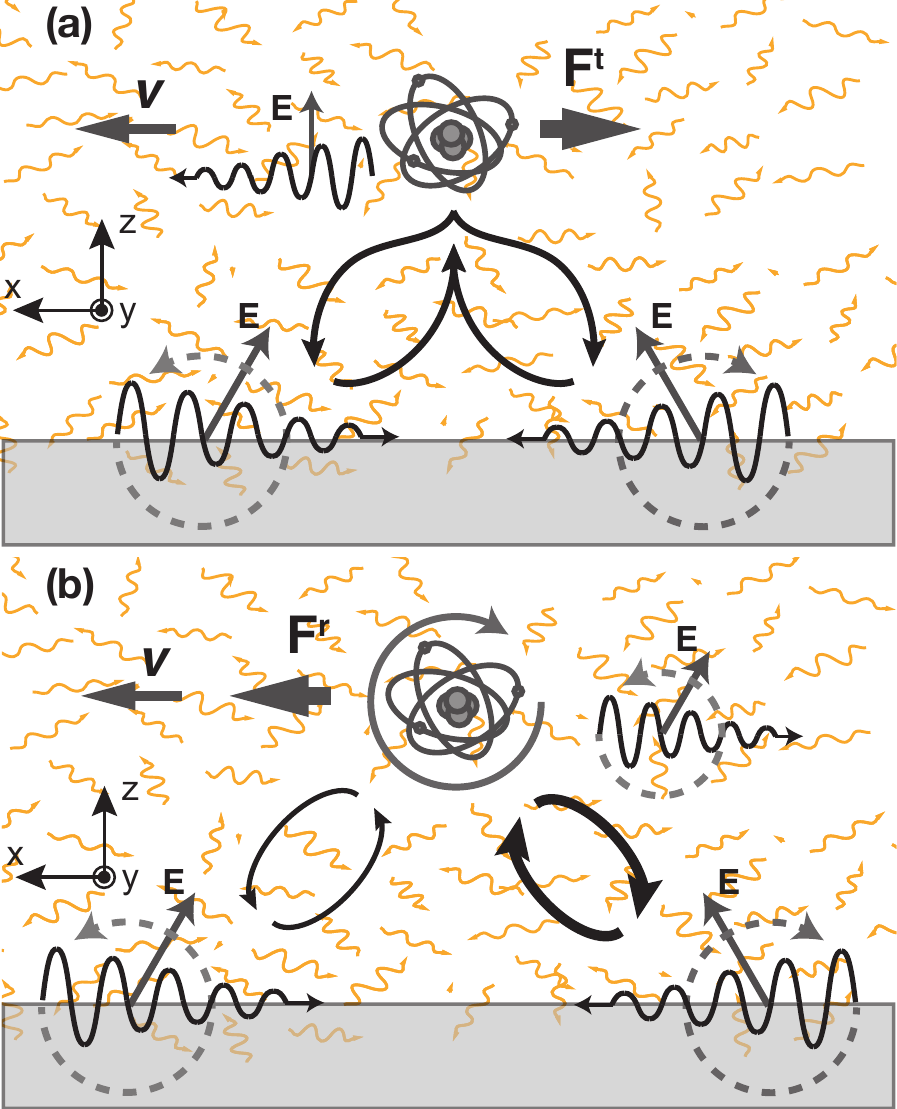}
\caption{
Schematic description of the two mechanisms behind Eqs.~\eqref{friction2total}.
{\bf (a)} In the near field two counter-propagating virtual surface excitations 
give rise to spin-zero photons (corresponding to a linearly polarized electromagnetic 
field $\mathbf{E}$) that are absorbed by the atom which gets excited (anomalous 
Doppler effect). The atom then emits linearly polarized photons, predominantly 
in the direction of the motion. The corresponding recoil momentum gives rise to 
$\mathbf{F}^{\rm t}$. {\bf (b)} When the atomic rotational degrees of freedom are 
involved in the dynamics, the atom can also absorb photons with nonzero spin
(corresponding to circularly polarized electromagnetic field $\mathbf{E}$). 
For a motion along the positive $x$-axis, surface excitations with negative 
spin are predominantly absorbed, producing a clockwise rotation of the atom 
around the $y$-axis. In this case, the atom prevalently emits photons with 
nonzero spin in the negative $x$-direction, corresponding to the recoil 
force $\mathbf{F}^{\rm r}$ oriented in the direction of the motion.}
\label{RollFricProcessesFig}
\end{figure}

The two terms in Eqs.~\eqref{friction2total} correspond to two distinct physical 
mechanisms characterizing the system [see Fig.~\ref{RollFricProcessesFig}].
A first insight about their origin is provided by looking at the correlation tensor:
If $\underline{C}(\tau,\mathbf{v})
=\underline{C}^{\sf T}(\tau, \mathbf{v})$, $\underline{S}(\omega,\mathbf{v})$ 
is necessarily real and symmetric, leading to $\mathbf{F}^{\rm t}\not=0$ and 
$\mathbf{F}^{\rm r}=0$ \cite{Intravaia14,Intravaia16}. This condition is 
equivalent to $\langle\mathbfh{d}(\tau)\times\mathbfh{d}(0)\rangle=0$, which 
implies that, on average, the atomic dipole cannot rotate, absorb or emit any 
net angular momentum. $\mathbf{F}^{\rm t}$ is therefore the quantum frictional 
force commonly investigated in the literature, which only takes into account 
the atomic translational motion. $\mathbf{F}^{\rm r}$ thus represents an additional 
contribution which appears if the rotational atomic degrees of freedom are 
considered and is the main focus of this work.

In order to obtain a deeper understanding, it is useful to analyze the Green tensor 
of our system. Without loss of generality we consider a motion along the $x$-direction 
($\mathbf{v}=v \mathbf{e}_{x}, |\mathbf{e}_{x}|=1$). The surface-related (scattering) part of
$\underline{G}(\mathbf{k},z_a,\omega)$ can then be written as the sum of a diagonal 
and a skew-symmetric matrix, $\underline{\sigma}(\mathbf{k},z_{a}, \omega)$ and 
$-\phi(\mathbf{k},z_{a}, \omega)\underline{L}_{y}$ respectively. $\underline{L}_{y}$ 
is the $y$-component of the usual Lie-algebra's basis for {\bf so}(3) 
($[\underline{L}_{i}]_{jk}=-\imath \varepsilon_{ijk}$) describing 3D-rotations 
\cite{Note1}.
As we will see in detail below [see Eqs.~\eqref{SGJ}], $\underline{G}(\mathbf{k},z_a,\omega)$, 
describing the system's electromagnetic response, and $\underline{S}(\omega,\mathbf{v})$, 
accounting for the atomic fluctuations, are in general physically connected. The link 
is provided by the matrix 
$\underline{G}_{\Im}(\mathbf{k},z_{a}, \omega)=[\underline{G}(\mathbf{k},z_{a}, \omega)-\underline{G}^{\dag}(\mathbf{k},z_{a}, \omega)]/(2\imath)=\underline{G}^{\rm s}_{I}(\mathbf{k},z_{a},\omega)-\imath\underline{G}^{\rm as}_{R}(\mathbf{k},z_{a},\omega)$, which is related to the probability that the atom absorbs ($\omega<0$) or emits ($\omega>0$) photons 
\cite{Novotny06}.
With reference to the electromagnetic spin operator 
\cite{Aiello15,Bliokh15,Mandel95}, 
the structure inherited from the Green tensor reveals that the interaction is 
sensitive to the three states of the photon's spin.  
$\underline{G}^{\rm s}_{I}(\mathbf{k},z_{a},\omega)=\underline{\sigma}_{I}(\mathbf{k},z_{a}, \omega)$ 
is associated with linearly polarized photons (spin zero): Due to the matrix' even 
parity in $\mathbf{k}$, the corresponding processes do not depend on the direction 
of propagation. 
In contrast, the matrix $\phi_{I}(\mathbf{k},z_{a}, \omega)\underline{L}_{y}=\underline{G}^{\rm as}_{R}(\mathbf{k},z_{a},\omega)$, 
describes emission and/or absorption of photons having a nonzero spin 
along the $y$-axis. 
The interpretation in terms of absorption and emission probability implies a 
positive spin when $\phi_{I}<0$ and a negative spin in the opposite case, 
linking the sign to the direction of propagation through the odd parity in 
$\mathbf{k}$ of the function $\phi$. 
This locking behavior, which is essentially associated with the confinement of light 
at the vacuum-material interface 
\cite{Aiello15,Bliokh15,Lodahl17}, 
allows to associate $\phi_{I}(\mathbf{k},z_{a}, \omega)$ with a spin-dependent local density of states. 
In the near-field limit we have \cite{Wylie85,Intravaia16a}
\begin{subequations}
\label{Gdef}
\begin{gather}
\underline{G}(\mathbf{k},z_{a}, \omega)\approx
			\underline{\Pi}\,\frac{k}{2\epsilon_{0}} 
						r[\omega]e^{-2k z_{a}}\quad (k=|\mathbf{k}|),
\label{retardedG}
\\
\underline{\Pi}
=
\begin{pmatrix}
\frac{k_{x}^{2}}{k^{2}}&\frac{k_{y}k_{x}}{k^{2}}&-\imath \frac{k_{x}}{k}\\
		\frac{k_{y}k_{x}}{k^{2}}&\frac{k_{y}^{2}}{k^{2}}&-\imath \frac{k_{y}}{k}\\
								\imath \frac{k_{x}}{k}&\imath \frac{k_{y}}{k}&1
\end{pmatrix} 
\to
\begin{pmatrix}
\frac{k_{x}^{2}}{k^{2}}&0&-\imath \frac{k_{x}}{k}\\
		0&\frac{k_{y}^{2}}{k^{2}}&0\\
								\imath \frac{k_{x}}{k}&0&1
\end{pmatrix}.
\end{gather}
\end{subequations}
$\epsilon_{0}$ is the vacuum permittivity and 
$r[\omega]$ the surface's p-polarized reflection coefficient. 
Please note, that in the r.h.s. expression we have deleted those terms 
that -- due to symmetry -- do not contribute to Eq.~\eqref{friction2total}. 
As $\underline{\Pi}^{\dag}=\underline{\Pi}$, $\underline{G}_{\Im}$ is obtained 
by replacing $r[\omega]$ with $r_{I}[\omega]$ in Eq.~\eqref{retardedG}. 
At equilibrium and for common materials
the radiation features zero angular momentum on average.  However, for a moving 
atom, the frequency of the radiation in the co-moving frame is Doppler-shifted 
by the value $\mathbf{k}\cdot\mathbf{v}$ [Eqs.~\eqref{friction2total}]. This induces 
an asymmetry in the spin balance and the atom effectively
perceives spin-polarized radiation.

\begin{figure}
\includegraphics[width=0.49\textwidth]{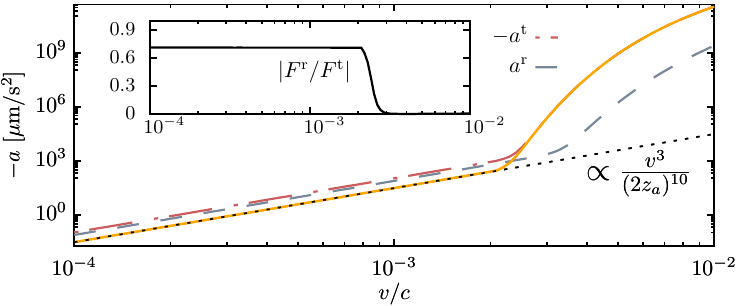}
\caption{Frictional acceleration, $a=a^{\rm t}+a^{\rm r}$, on a $^{87}$Rb atom 
($\alpha_{0}= 4\pi \epsilon_{0} \times 47.28$ \AA$^{3}$, $\omega_{a}=1.3$ eV, 
$m_{\mathrm{Rb}}=86.9$ u \cite{Steck08}) as a function of its velocity. 
The particle moves at $z_{a}=5$ nm from a gold surface, described by a Drude-permittivity 
$\epsilon(\omega)=1-\omega_{\rm p}^{2}/[\omega(\omega+\imath \Gamma)]^{-1}$ 
($\omega_{\rm p}=9$ eV, $\Gamma=35$ meV \cite{Barchiesi14}, giving 
$\rho=\Gamma/[\epsilon_{0}\omega_{\rm p}^{2}]=3.21\times 10^{-8}$ $\Omega$m). 
The two competing contributions $a^{\rm t}$ (dash-dotted line) and 
$a^{\rm r}$ (dashed line) in Eqs.~\eqref{friction2total} are represented. 
At small velocity the total acceleration (full line) scales as the sum of 
the expressions in Eq.~\eqref{asymp1} divided by the atomic mass (dotted line), 
while at high velocity friction is enhanced by a resonant interaction 
\cite{Intravaia16}.
The \textit{inset} shows that at low velocities $F^{\rm r}$ compensates more 
than 70\% of $F^{\rm t}$. The percentage decreases at higher velocity.}
\label{FricForce}
\end{figure}

To quantitatively understand the implications of this phenomenon on the frictional 
force, we model the atom's internal structure as a Lorentz harmonic oscillator 
characterized by the transition frequency $\omega_{a}$. 
The velocity-dependent atomic polarizability tensor is then given by 
\cite{Intravaia16}
\begin{equation}
\underline{\alpha}(\omega,\mathbf{v})=
					\alpha_{B}(\omega)
						\left[1 -\alpha_{B}(\omega) \int \frac{d^2{\bf k}}{(2 \pi)^{2}}\underline{G}({\bf k},z_{a},\omega^{+}_{\mathbf{k}}) \right]^{-1},
\label{alphaV}
\end{equation}
where $\alpha_{B}(\omega)=\alpha_{0}\omega_{a}^{2}/(\omega^{2}_{a}-\omega^{2})$ and 
$\alpha_{0}$ 
are the bare and static oscillator's polarizabilities, respectively. 
The nonequilibrium power spectrum can be written as 
\begin{subequations}
\label{SGJ}
\begin{equation} 
\underline{S}(\omega,\mathbf{v}) =
							\frac{\hbar}{\pi}  \left[\theta(\omega)
								\underline{\alpha}_{\Im}(\omega,\mathbf{v}) +  \underline{J}(\omega, \mathbf{v})\right],
\label{non-eq-FDT}
\end{equation}
 where $\underline{\alpha}_{\Im}(\omega,\mathbf{v})$ is defined similarly to $\underline{G}_{\Im}(\mathbf{k},z,\omega)$  and
\begin{align}
\underline{J}(\omega,\mathbf{v}) = \int &\frac{d^2{\bf k}}{(2\pi)^2} 
								[\theta(\omega^{+}_{\mathbf{k}}) - \theta(\omega)]  \nonumber\\
									&\times \underline{\alpha}(\omega,\mathbf{v}) 
										 \cdot  \underline{G}_{\Im}({\bf k},z_{a},\omega^{+}_{\mathbf{k}}) 
											\cdot \underline{\alpha}^{\dag}(\omega,\mathbf{v}).
\end{align}
\end{subequations}
The nonequilibrium fluctuation theorem in Eqs.~\eqref{SGJ} includes the atomic 
rotational degrees of freedom and generalizes results reported in previous work \cite{Intravaia16}.
Equations \eqref{alphaV} and \eqref{SGJ} can be used to evaluate Eqs.~\eqref{friction2total}. 
For illustration, we present the resulting frictional deceleration on a $^{87}$Rb atom 
moving  above a gold surface in Fig.~\ref{FricForce}. For symmetry reasons 
the deceleration is along the direction of motion. Notice that the positive rotational 
contribution attenuates the frictional force stemming from the translation. Roughly 
speaking, one can say that, as in classical mechanics, allowing for a ``rolling dynamics'' 
reduces the frictional force acting on the object.

For additional insight and a more quantitative analysis, we focus on the low-velocity 
limit of Eqs.~\eqref{friction2total}. 
As discussed in previous work \cite{Intravaia14,Intravaia16,Intravaia16a}, the dominant 
contribution to the frictional force arises from low frequencies $\omega\lesssim v/z_{a}$. 
If $r_{I}\approx 2 \epsilon_{0}\rho \omega$ for these frequencies (for conductors 
$\rho$ is the resistivity) \cite{Intravaia16,Reiche17,Oelschlager18}, to second order 
in $\alpha_{0}$,  we have
\begin{subequations}
\label{RollingFricLowV}
\begin{multline}
F^{\rm t}\approx -\alpha^{2}_{0} v^{3}\frac{\hbar}{\pi} 
				\int\frac{d^{2}\mathbf{k}}{(2\pi)^{2}}\frac{d^{2}\tilde{\mathbf{k}}}{(2\pi)^{2}}
							\frac{k_{x}}{6}(k_{x}+\tilde{k}_{x})^{3}
								\\
						\times\mathrm{Tr}\left[\underline{\sigma}'_{I}(\mathbf{k},z_{a},0)
								\cdot\underline{\sigma}'_{I}(\tilde{\mathbf{k}},z_{a},0)\right],
\label{Ftlowv}
\end{multline}
\begin{multline}
F^{\rm r}\approx -\alpha^{2}_{0}v^{3}\frac{\hbar}{\pi} 
				\int\frac{d^{2}\mathbf{k}}{(2\pi)^{2}} \frac{d^{2}\tilde{\mathbf{k}}}{(2\pi)^{2}}
							\frac{k_{x}}{6}(k_{x}+\tilde{k}_{x})^{3}
								\\
						\times\mathrm{Tr}[\underline{L}^{\sf T}_{y}\cdot\underline{L}_{y}]
									\phi'_{I}(\mathbf{k},z_{a},0)\phi'_{I}(\tilde{\mathbf{k}},z_{a},0),
\label{Frlowv}
\end{multline}
\end{subequations}
where the prime indicates the derivative with respect to the frequency.
For a motion within the near field of the surface, Eqs.~\eqref{RollingFricLowV} give
\begin{equation}
F^{\rm t}\approx -\frac{63}{\pi^{3}} \hbar\alpha_{0}^{2}\rho^{2}
 					\frac{v^{3}}{(2 z_a)^{10}},\quad 
F^{\rm r}\approx \frac{45}{\pi^{3}}\hbar\alpha_{0}^{2}\rho^{2}
					\frac{v^{3} }{(2z_a)^{10}}~.
\label{asymp1}
\end{equation}
Notice that the contribution associated with the rotation compensates more than 70\% 
of the force related to the translation (see inset in in Fig.~\ref{FricForce}).
Interestingly, one can show that at low velocity, using the so-called local thermal 
equilibrium (LTE) approximation, the compensation between the translational and the 
rotational contributions is complete, leading to an erroneous vanishing frictional 
force. The LTE approach is commonly used for an approximate description of 
nonequilibrium systems and treats each of its components as if they were locally in 
thermal equilibrium with their immediate surrounding. The (equilibrium) 
fluctuation-dissipation theorem (FDT) is then applied 
\cite{Callen51}, 
which for our system is equivalent to neglecting $\underline{J}$ in Eq.~\eqref{non-eq-FDT}. 
A vanishing friction in the LTE approximation indicates that the detailed balance 
enforced by the FDT incorrectly treats the processes connected with the translation 
and the rotation on the same footing.
Contrasted with Eq.~\eqref{asymp1}, the LTE result is not only flawed but also 
highlights that, as soon as the rotational degrees of freedom are included, quantum 
friction is essentially a pure nonequilibrium phenomenon.

The difference in sign between $F^{\rm t}$ and $F^{\rm r}$ can be understood 
as a consequence of the interplay between the anomalous Doppler effect and 
the spin-momentum locking of light. In the NESS, for a motion along $x>0$, 
the Doppler-shifted frequency $\omega^{-}_{\mathbf{k}}$ becomes ``anomalously'' 
negative only for $k_{x}>0$. Therefore, during the motion the atom can get 
excited even at zero temperature [see the discussion before Eq.~\eqref{Gdef}] 
due to a light-matter interaction that favors positive $k_{x}$.  
In Eq.~\eqref{Ftlowv} emission and absorption are controlled by 
$\underline{\sigma}(\mathbf{k},z_{a}, \omega)$ and involve spin-zero photons 
which are prevalently emitted in the direction of the motion. As a result,
 the net recoil force $F^{\rm t}$ acts against the motion \cite{Intravaia14,Intravaia16,Intravaia16a}. 
However, allowing for the atomic rotational dynamics opens an additional channel 
of interaction with the surface represented by Eq.~\eqref{friction2r}. The 
corresponding processes in Eq.~\eqref{Frlowv} are associated with the product 
$\phi(\mathbf{k},z_{a}, \omega)\underline{L}_{y}$ and involve photons with 
nonzero spin. In this case, the angular momentum acts as an additional filter 
that selects a prevalent emission along the negative $x$-direction. This leads 
to the net recoil force $F^{\rm r}$ with the same sign of the velocity.

The involvement of the angular momentum has an additional implication. During 
the motion and the interaction with the vacuum field, the atom undergoes a 
stochastic process which includes rotation.
The stationarity characterizing the NESS implies that all torques acting on the 
atom, resulting from the dissipative nonequilibrium light-matter interaction, 
must balance on average. 
The rotational stochastic motion \cite{Stickler18} generated by the exchange of 
photons can be associated with a constant angular momentum 
$\boldsymbol{\mathcal{L}}=[\alpha_{0}\omega_{a}^{2}]^{-1}\langle \mathbfh{d}(t)\times \dot{\mathbfh{d}}(t)\rangle$, 
which can be written as 
\cite{Dennis04}
\begin{equation}
\label{Angular}
\boldsymbol{\mathcal{L}}=\frac{1}{\alpha_{0}\omega_{a}^{2}}
					\int_{-\infty}^{\infty}d\omega~
								\omega\mathrm{Tr}[\underline{\mathbf{L}}\cdot\underline{S}(\omega,\mathbf{v})].
\end{equation}
In agreement with the symmetries of our system, only the $y$-component is nonzero. 
We can evaluate the corresponding rotation frequency $\Omega$ by multiplying the 
angular momentum by the inverse of the average atomic moment of inertia tensor, 
$\underline{M}_{ij}=[\alpha_{0}\omega_{a}^{2}]^{-1}\langle|\mathbfh{d}(t)|^{2}\delta_{ij}-\hat{d}_{i}(t)\hat{d}_{j}(t)\rangle$. 
With some matrix algebra this yields
\begin{align}
\label{RotFreq}
\Omega=\frac{\int_{-\infty}^{\infty}d\omega ~  
				\omega\mathrm{Tr}[ \underline{S}(\omega,\mathbf{v}) \cdot\underline{L}_{y}]}
						{\int_{-\infty}^{\infty}d\omega\, \mathrm{Tr}[\underline{S}(\omega,\mathbf{v}) \cdot\underline{L}^{2}_{y}]}~.
\end{align}
Inserting Eq.~\eqref{non-eq-FDT} in Eq.~\eqref{RotFreq}, in the near-field limit 
and for $\omega_{a}$ within the Ohmic response of the material, we obtain at the 
leading order in $\alpha_{0}$
\cite{Note2}
\begin{equation}
\Omega
\approx 
-\frac{v}{z_{a}}\left[1+\left(\frac{1}{3}\frac{r_{R}[\omega_{a}]}{r_{I}[\omega_{a}]}\right)^{2}\right]^{-1}~.
\label{RotFreqV}
\end{equation}
Equation \eqref{RotFreqV} indicates that, in the NESS, while propelled by a constant 
external force near the surface, translation and rotation couple and the atom rotates 
\emph{clockwise} around the $y$-axis despite no external torque is applied with respect 
to an axis passing through its center of mass (see Fig.~\ref{RollFricProcessesFig}).
This last result contradicts our classical intuition, which, for a motion along the 
positive $x$-axis, would instead suggest a counterclockwise rotation. It also differs 
from evaluations of purely rotating metallic nanoparticles without translational motion
\cite{Zhao12,Rodriguez-Fortuno15,Manjavacas17,Dedkov17}, 
or on immobile circularly polarized excited atoms 
\cite{Oude-Weernink18} 
in front of a surface, which in the near-field and low rotational frequency limits predict 
lateral forces agreeing with the classical prescription.
Once again, however, the sense of rotation can be interpreted as resulting from the 
motion-induced asymmetry in the light-matter interaction.
In the atomic excitation process the anomalous Doppler effect favors the absorption 
of photons propagating along the positive $x$-axis. In the near-field they have negative 
spin, resulting in the absorption of negative angular momentum and a clockwise rotation 
of the atom. This also provides a better understanding of the sign of $F^{\rm r}$: During 
the dissipative process associated with the frictional force, in order to keep 
$\boldsymbol{\mathcal{L}}$ constant, the atom emits photons with positive spin, thus 
absorbing a negative angular momentum recoil. Due to the properties of the spin-dependent 
density of states, in the near field these photons can be absorbed by the environment 
(essentially the surface) if they are emitted along the negative $x$-axis, thus favoring 
a positive momentum recoil and a positive $F^{\rm r}$. Still, because of the Doppler-shift, 
the last process is less effective than the one associated with spin-zero photons, whose 
absorption rate does not depend on the sign of the wave vector, justifying why 
$\abs{F^{\rm r}}<\abs{F^{\rm t}}$.

It is important to highlight that our description takes into account the full nonequilibrium 
electromagnetic backaction on the microscopic object, setting it apart from other related 
studies. 
Nonequilibrium backaction is often not included in perturbative approaches for atoms 
\cite{Scheel15,Oude-Weernink18} 
and it is commonly neglected for metallic nanoparticles 
\cite{Zhao12,Manjavacas17,Dedkov17}, 
due to the strong intrinsic dissipation of the metal 
\cite{Intravaia16a}.
In our case, however, this feature ultimately characterizes important quantities such 
as the atomic power spectrum or polarizability and affects the spin-sensitive atom-surface 
interaction.  Disregarding the backaction removes the intrinsic velocity-dependence in 
these quantities, making them coincide with their bare or equilibrium expressions. This 
leads to a description which, to a large extent, is equivalent to the LTE approximation 
for which some of the above effects disappear.

The measurement of the quantum frictional force and of the corresponding deceleration 
is challenging due to the weakness of the interaction. It requires a careful choice 
of both the experimental technique and the system's parameters.
From Fig.~\ref{FricForce}, we see that on a Rubidium atom moving at $v\gtrsim 30$ km/s 
at a distance of 5 nm from a Gold surface acts a deceleration 
$|a|\gtrsim 3 \times 10^{-2}$ $\mu$m/s$^{2}$. Notice, however, that replacing 
$^{87}$Rb with $^{7}$Li ($\alpha_{0}=4\pi \epsilon_{0} \times 24.33$ \AA$^{3}$ 
\cite{Miffre06}, 
$m_{\mathrm{Li}}=7.02$ u) and Gold with Sodium ($\rho= 8\times 10^{-7}$ $\Omega$m 
\cite{Zeman87}), 
already leads to a deceleration of $2.5$ $\mu$m/s$^{2}$ for $v\approx 10$ km/s. 
In addition, previous work has indicated that considering spatial dispersion and 
engineering the surface's optical response can enhance the effect by several 
orders of magnitude 
\cite{Reiche17,Oelschlager18}. 
Consequently, promising perspectives for a detection are offered by atom interferometry. 
Cold-atom setups have already achieved an accuracy of $ 10^{-2}$ $\mu$m/s$^{2}$ 
\cite{Cronin09} for the measurement of accelerations. 
To increase the (relative) velocity, one can consider atoms moving close to or trapped in a 
ring-shaped potential \cite{Navez16} parallel to a 
surface rotating with high frequency ($v\sim10$ km/s are reached in micro-turbines). Alternatively, 
$v\gtrsim 100$ km/s are achieved using neutralized ion beams \cite{Brand19}. In this 
last case, one can search for friction-induced modifications of the interference 
pattern produced by the diffraction of the atomic beam on a grating with very small 
apertures ($\sim$ 45-50 nm 
\cite{Perreault05a,Brand15,Note3}
).
An indirect proof for $F^{\rm r}$ can be provided by the detection of the atomic 
rotation. 
Using the parameters of Fig.~\ref{FricForce}, for $v\approx10$ km/s we obtain 
$|\Omega|\approx 25$ MHz, which can be detected by measuring the atomic optical 
response to circular polarized light.

In conclusion, we have shown that the combination of nonequilibrium dynamics, the 
anomalous Doppler effect and the spin-momentum locking of light induces quantum 
rolling friction on an atom moving parallel to a surface. 
During the zero-temperature dissipative process, the atom performs a driven 
Brownian-like motion that involves its internal degrees of freedom and depends 
on the three states of the photon's spin.
The atom absorbs and emits photons, exchanging translational and angular momentum
with light. 
As in classical rolling motion, the interplay between atomic translational and 
rotational motion sensibly diminishes the drag force with respect to the case 
where the rotation is not considered. 
Interestingly, however, the reduction in strength of friction is connected 
with a steady atomic rotation with a sense opposite to what one would expect from 
classical intuition. 
Our analysis qualitatively applies to a large class of systems and materials 
showing similar features (low-frequency dissipation, light confinement etc.). 
It also suggests ways for tuning the total quantum frictional interaction via 
an enhancement of the system's asymmetry.  They can involve for instance, chiral 
atoms 
\cite{Barcellona17}, 
topological materials 
\cite{Hassani-Gangaraj18}, 
or even external fields 
\cite{Rebane12}, 
which can affect the exchange of angular momentum within the system.
Quantum rolling friction yields an interesting example on how different phenomena 
unconventionally combine in the quantum realm.

{\it Acknowledgments.} We thank R. O. Behunin, A. Manjavacas, P. Schneeweiss, 
A. Rauschenbeutel and R. Decca for insightful discussions. We acknowledge funding by the 
Deutsche Forschungsgemeinschaft (DFG) through SFB 951 HIOS (B10, Project No. 182087777)
and by the LANL LDRD program. FI further acknowledges financial support 
from the DFG through the DIP program (grant FO 703/2-1 and SCHM 1049/7-1)


\end{document}